\documentclass{article}
\usepackage{bm} % Math packages
\usepackage{bbm}
\usepackage{amsmath}
\usepackage[flushleft]{threeparttable}
\usepackage{booktabs}
\usepackage{amssymb}
\usepackage[ruled,vlined,linesnumbered,noresetcount]{algorithm2e}
\usepackage{caption}
\usepackage{subcaption}
\usepackage{adjustbox}
\usepackage[utf8]{inputenc}
\usepackage[english]{babel}
\usepackage{graphicx}
\usepackage{array}
\usepackage{verbatim}
\usepackage{float}

\usepackage[
backend=biber,
style=authoryear,
doi=false,
url=false,
eprint=false,
]{biblatex} %Imports biblatex package
\usepackage[a4paper,top=3cm,bottom=2cm,left=3cm,right=3cm,marginparwidth=1.75cm]{geometry}
\usepackage{graphicx}
\graphicspath{ {images/} }

\newtheorem{assumption}{Assumption}
\newtheorem{lemma}{Lemma}
\newtheorem{theorem}{Theorem}
\begin{document}

\title{Predictor Selection for Synthetic Controls}
%
%\titlerunning{SSC}  % abbreviated title (for running head)
%                                         also used for the TOC unless
%                                     \toctitle is used
%

\author{Jaume Vives-i-Bastida\thanks{Department of Economics, Massachusetts Institute of Technology. Email: vives@mit.edu. I thank Alberto Abadie, Amy Finkelstein, Anna Mikusheva, Stephen Morris, Brian Quistorff, Jason Thorpe and the participants in the MIT Economics second year research seminar, the MIT econometrics lunch, the Google statistics seminar, and the Meta economics seminar for valuable comments and suggestions. An earlier version of the paper was circulated under the title \textit{Sparse Synthetic Controls}.}}
% "Quistorff, Brian"
%\authorrunning{Jaume Vives} % abbreviated author list (for running head)
%
%%%% list of authors for the TOC (use if author list has to be modified)
%
%\institute{MIT}

\maketitle              % typeset the title of the contribution

\begin{abstract}
Synthetic control methods often rely on matching pre-treatment characteristics (called predictors) of the treated unit. The choice of predictors and how they are weighted plays a key role in the performance and interpretability of synthetic control estimators. This paper proposes the use of a \textit{sparse} synthetic control procedure that penalizes the number of predictors used in generating the counterfactual to select the most important predictors. We derive, in a linear factor model framework, a new model selection consistency result and show that the penalized procedure has a faster mean squared error convergence rate. Through a simulation study, we then show that the sparse synthetic control achieves lower bias and has better post-treatment performance than the un-penalized synthetic control. Finally, we apply the method to revisit the study of the passage of Proposition 99 in California in an augmented setting with a large number of predictors available.

\noindent \textbf{keywords}: synthetic controls, lasso, variable selection, linear factor models, high-dimensional observational studies.
\end{abstract}

%\begin{center}
%	[PRELIMINARY DRAFT]
%\end{center}
%
\section{Introduction}
Synthetic controls have become a popular method for making inferences on causal effects of policy interventions. Its applications have ranged from evaluating the effects of tax changes (\textcite{adh2010}, \textcite{kleven}) to more complex policy changes such as guaranteed job programs (\textcite{kasy}). Beyond the social sciences synthetic controls have also been used in other applied sciences, for example in engineering and the health sciences, and in the private sector, for instance to evaluate the effect of advertising promotions in large tech companies. 

In the classical synthetic control setting (\textcite{abadie2003}) a single aggregate unit (such as a city or a state) is exposed to a policy treatment at period $T_0$, and $J$ units that are never exposed to the policy (called the donor units) are used to generate a counterfactual. The researcher builds the synthetic control by finding the combination of donor units that best matches the pre-treatment characteristics (called the predictors) of the treated unit. Predictors may include both pre-treament outcomes as well as other covariates that are informative of the outcome of interest. An open question in the literature regards best practices and theory on how to choose the set of predictors and how to weight the importance of each predictor in building the synthetic control. 

There are two main reasons why the choice of predictor weights matters for computing synthetic controls. First, as noted by \textcite{adh2015}, \textcite{abadievives}, \textcite{klosner} and \textcite{malo} among others, the predictor weights are important for the performance of the synthetic controls. Weighting the predictors naively can lead to bad pre-intervention fit and synthetic controls with poor post-intervention performance. This, in turn, means that the estimated treatment effects will be biased. Second, the predictor weights may be of interest in their own right. The researcher may be interested in interpreting them to explain how the synthetic control was built. For example, in the California tobacco control program application that we will re-visit in Section 5, the researcher may be interested in explaining what predictors (alcohol consumption, income etc) are used in generating the counterfactual. Hence, the choice of predictors is important both for the performance and the interpretability of the synthetic control.

Most of the literature relies on treating the predictor weights as a hyper-parameter and choosing them through a cross-validation procedure. Different implementations and best practices for choosing the weights have been proposed, see for example \textcite{adh2010}, \textcite{adh2015}, \textcite{doudchenko}, and \textcite{benmichael}. A common thread throughout, however, is that researchers should have a good prior on what predictors to include to construct the synthetic control. Recently, \textcite{pouliot2022} note that cross-validation may not be amenable to many synthetic control settings and insightfully propose an information criteria for model selection of the donor weights based on estimating the degrees of freedom of the synthetic control problem. Under a gaussianity assumption they show that using the information criteria leads to the right selection of the combination of donor units.

In this paper our focus is slightly different. We are interested in the selection of the predictors themselves as well as how they relate to the performance of the synthetic control. Taking \textcite{adh2015} as a starting point, we show theoretically and empirically that a modified penalized procedure that includes an $l_1$ penalty for the predictor weights can achieve consistent predictor selection as well as better mean squared error performance than the standard method. Our theoretical results for a linear factor model ensure that, almost surely, predictors that do not contribute to the outcome of interest in the model will not be used in the match to compute the synthetic control. Under stronger assumptions, the result also holds the other way around; predictors that are useful in generating the outcome of interest are used with positive probability. In the same setting we also derive a synthetic control bias bound akin to the one in \textcite{adh2010} and \textcite{fermanpinto}, and show that the bias depends on whether the useful predictors can be matched well. Finally, we show that the proposed penalized procedure has a faster mean square error rate of convergence for matching the predictors, suggesting it may have lower bias than the standard method.

We refer to our proposed procedure as the \textit{sparse} synthetic control as it is related to the class of penalized synthetic control models proposed by \textcite{sparseSC}. The focus of \textcite{sparseSC}, however, is on showing consistency and inference for penalized synthetic control estimators of the average treatment effect on the treated using the theoretical framework of \textcite{ferman} and \textcite{chernsc}. Our model consistency results also apply to the estimators proposed by \textcite{sparseSC}, therefore we see our theoretical results and proposed methodology as complementary to \textcite{sparseSC} in the analysis of the properties of penalized synthetic control methods. 

To evaluate the usefulness of the theoretical results in practice we compare in a simulation study the standard (un-penalized) synthetic control and our proposed method and revisit the passage of Proposition 99 in California  using an extended data set with forty predictors. The simulation evidence confirms that the penalized method is able to identify the useful from the nuissance predictors and achieves better mean squared error in the post-treatment period. The empirical application highlights that when the number of predictors is large the standard un-penalized synthetic control may struggle to provide a valid synthetic control. On the other hand, our procedure is robust to increasing the number of predictors from the original seven used in \textcite{adh2010} to a higher dimensional setting with forty predictors. Furthermore, the \textit{sparse} synthetic control uses predominantly predictors included in the original application, confirming the predictor choice of \textcite{adh2010}. Finally, an estimate of the placebo variance in the application suggests that the proposed method has lower variance than the standard method, as predicted by the theory.

%This paper draws from the classic synthetic control literature Abadie and Gardeazabal 2003 and Abadie, Diamond, and Hainmueller (2010, 2015), and the growing literature on extensions to the synthetic control method. Most notably, this paper is relevant for two strains of the literature. First, it complements the literature on penalized methods for synthetic controls (Abadie and L'Hour 2016, Doudchenko and Imbens 2016, Ben-Michael, Feller, and Rothstein 2018, and Arkhangelsky, Athey, Hirshberg,  Imbens and Wager 2020) by focusing on penalizing the predictor weights rather than the donor unit weights. In doing so, I show that using a penalized method to choose the predictor weights can improve performance. Second, it complements the literature on how to choose the predictor weights (Klosner, Kaul, Pfeifer and Schieler 2018 and Malo, Eskelinen, Zhou and Kuosmanen 2020) by providing a new methodology and a new model selection result. Overall, this paper addresses a new question not previously considered in the literature which is how to consistently select the predictors used in building the synthetic control and provides a method to do it with good performance properties.

The paper is structured as follows. Section 2 describes the proposed penalized synthetic control method and a computation algorithm. Section 3 presents the main theoretical results in a linear factor model. Section 4 explains the simulation study results, and finally Section 5 discusses the empirical application to the California tobacco control program.

\section{Sparse Synthetic Controls}
 
To define the \textit{sparse} synthetic control method, consider a setting in which we observe $J+1$ aggregate units for $T$ periods. The outcome of interest is denoted by $Y_{it}$ and only unit 1 is exposed to the intervention during periods $T_0+1, \dots, T$. We are interested in estimating the treatment effect $\tau_{1t} = Y_{1t}^I - Y_{1t}^N$ for $t>T_0$, where $Y_{1t}^I$ and $Y_{1t}^N$ denote the outcomes under the intervention and in absence of the intervention respectively. Since we do not observe $Y_{1t}^N$ for $t>T_0$ we estimate $\tau_{1t}$ by building a counterfactual $\hat{Y}_{1t}^N$ of the treated unit's outcome in absence of the intervention.

As in the standard synthetic control our counterfactual outcome will be given by a weighted average of the donor units' outcomes, that is  $\hat{Y}_{1t}^N = \sum_{j=2}^{J+1} w_jY_{j t}$ for a set of weights $\mathbf{w} = (w_2, \dots, w_{J+1})'$. To choose the weight vector $\mathbf{w}$ we use observed characteristics of the units and pre-intervention measures of the outcome of interest. Formally, we let the $K \times 1$ design matrix for the treated unit be $\mathbf{X}_1 = (Z_1, \bar{Y}_1^{\mathbf{K}_1}, \dots, \bar{Y}_1^{\mathbf{K}_M})'$, where  $\{\bar{Y}_1^{\mathbf{K}_i}\}_1^M$ represent $M$ linear combination of the outcome of interest for the pre-intervention period and $Z_1$ are pre-treament characteristics of the treated unit. Similarly, for the donor units, $\mathbf{X}_0$ is a $K\times J$ matrix constructed such that its $j$th column is given by  $(Z_j, \bar{Y}_j^{\mathbf{K}_1}, \dots, \bar{Y}_j^{\mathbf{K}_M})'$. We call the $K$ rows of the design matrices $\mathbf{X}_0$ and $\mathbf{X}_1$ the \textit{predictors} of the outcome of interest. This can include, for example, lags of the outcome variable and important context dependent characteristics of the aggregate units averaged over the pre-treatment period. 

We partition the pre-intervention period into a training set $(\mathbf{X}_0^{train},\mathbf{X}_1^{train}, \mathbf{Y}_0^{train}, \mathbf{Y}_1^{train})$ for $t\in \{1, \dots, T_v\}$  and a validation set $(\mathbf{X}_0^{val},\mathbf{X}_1^{val}, \mathbf{Y}_0^{val}, \mathbf{Y}_1^{val})$ for $t\in \{T_v +1, \dots, T_0$\}. This allows for the training and validation design matrices to differ, for example if predictors are averaged over different time periods or different linear combinations of lagged outcome variables are used.

The \textit{sparse} synthetic control is defined by the tuple of weight vectors $(\mathbf{V}^*, \mathbf{w}^*)$ computed by solving the following bi-level optimization program:\footnote{I follow the notation of Malo, Eskelinen, Zhou and Kuosmanen 2020, who propose a computational method to solve a similar problem.}

\begin{itemize}
    \item Upper level problem:
    \begin{align*}
	(\mathbf{V}^*, \mathbf{w}^*) \in \text{argmin}_{\mathbf{V},\mathbf{w}} &L_V(\mathbf{V},\mathbf{w}, \lambda) = \frac{1}{T_{val}} \|\mathbf{Y}_1^{val} - \mathbf{Y}_0^{val}\mathbf{w}(\mathbf{V})\|^2 + \lambda \| \text{diag}(\mathbf{V})\|_1, \\
	\text{s.t.  } &\mathbf{w}(\mathbf{V}) \in \psi(\mathbf{V}), \bm V \in \mathbb{R}^K_+.
    \end{align*}
    \item Lower level problem:
    $$
    \psi(\mathbf{V}) \equiv \text{argmin}_{\mathbf{w}\in \mathcal{W}} L_W(\mathbf{V},\mathbf{w}) = \|\mathbf{X}^{train}_1 - \mathbf{X}^{train}_0\mathbf{w}\|^2_V,
    $$
     where,
\begin{align*}
	\mathbf{w}\in \mathcal{W} &\equiv \left\{\mathbf{w}\in \mathbb{R}^J\text{ } |\text{ } \mathbf{1}'\mathbf{w} = 1, \text{ } w_j\geq 0, \text{ } j=2,\dots, J+1\right\}, \\
	\mathbf{V}\in \mathcal{V} &\equiv \left\{\mathbf{V} \text{ } | \text{ } \mathbf{V}\in \mathbb{R}^{K\times K}, \text{ } trace(\mathbf{V}) = 1, \text{ } V_{kk}\geq 0, V_{kl} = 0 \text{ for } k\neq l\right\},\text{ and} \\
	\| \cdot \|_V& \text{ denotes the semi-norm parameterized by $\mathbf{V}$ such that } \|\bm A\|_V = (\mathbf{A}'\mathbf{V}\mathbf{A})^{1/2}.
\end{align*}
\end{itemize}

The main idea of the \textit{sparse} synthetic control is that the $l_1$ penalty in the upper level problem induces some predictor weights (the diagonal elements of the weight matrix $\mathbf{V}$) to be set to zero as the penalty term $\lambda$ increases. A similar bi-level program is also considered in \textcite{sparseSC}, in which the authors also consider additional penalties for the lower level problem. In practice, given the weight restriction $\mathbf{V} \in \mathcal{V}$, we follow \textcite{adh2010} in the use of ex-post weight normalization by initially setting one predictor weight to one (i.e. $v_{k_0} = 1$ for some $k_0 \in \{1, \dots, K\}$) and only restricting the $v_k$ weights to be positive in the upper level program. In the appendix we describe why this ex-post normalization is necessary for finding a unique solution for the $v$ weights. The following algorithm details the procedure used to choose the hyper-parameter $\lambda$ and compute the \textit{sparse} synthetic controls.
\\

\begin{algorithm}[H]
\SetAlgoLined
\KwResult{$\mathbf{w}^*, \mathbf{V}^*$}
\KwData{$(\mathbf{X}_0^{train},\mathbf{X}_1^{train}, \mathbf{Y}_0^{train}, \mathbf{Y}_1^{train})$, $(\mathbf{X}_0^{train},\mathbf{X}_1^{train}, \mathbf{Y}_0^{val}, \mathbf{Y}_1^{val})$}
 set $v_{k_0}$= 1\;
 initialize $v_k$ for $k\neq k_0$ to $(\mathbf{X}_0^{train'}\mathbf{X}_0^{train})^{-1}$\;
 \For{each $\lambda$ in a grid}{
  get $(\mathbf{V}_{\lambda},\mathbf{w}_{\lambda})$ by jointly minimizing $L_W(\mathbf{V},\mathbf{w}, \lambda)$ and $L_V(\mathbf{V},\mathbf{w})$ for the training data\; 
  \hspace{ 1cm} s.t. $\mathbf{w}\in\mathcal{W}$, $v_k \geq 0$ $\forall k\neq k_0$ and $v_{k_0} = 1$\;
  scale $\mathbf{V_{\lambda}}$ to $[0,1]$\;
  get $\mathbf{w}^*_{\lambda}$ by minimizing $L_W(\mathbf{V_{\lambda}},\mathbf{w}, \lambda)$ for the training data\;
  store $\text{MSE}(\mathbf{Y}_1^{val}, \mathbf{Y}_0^{val}\mathbf{w}_{\lambda}^*)$ and $\mathbf{V}_{\lambda}$\;
 }
 choose $\lambda^*$ with minimum $\text{MSE}(\mathbf{Y}_1^{val}, \mathbf{Y}_0^{val}\mathbf{w}_{\lambda}^*)$\;
 $\mathbf{V}^* = \mathbf{V}_{\lambda^*}$\;
 get $\mathbf{w}^*$ by minimizing $L_V(\mathbf{V^*_{\lambda}},\mathbf{w})$ for the \textit{shifted} training data.\footnote{The shifted training data is the training data but with time dependent variables shifted to the $T_v$ periods before $T_0$.}
 \caption{\textit{Sparse} Synthetic Control}
\end{algorithm}

\section{Theoretical Results for a Linear Factor Model}

To motivate the model selection procedure for the\textit{sparse} synthetic controls theoretically, consider a standard setting in which the outcomes in absence of the intervention are given by a linear factor model as in \textcite{adh2010}

$$
Y_{it}^N = \delta_t + \bm{\theta}_t \mathbf{Z}_i + \bm{\lambda}_t \bm{\mu}_i + \epsilon_{it},
$$

\noindent 
where $\delta_t$ is a common factor with equal loadings, $Z_i$ is a ($k\times 1)$ vector of observed features, $\theta_t$ is a $(1\times k)$ vector of unknown parameters, $\lambda_t$ is a $(1\times F)$ vector of unobserved common factors, $\mu_i$ is an $(F\times 1)$ vector of unknown
factor loadings, and $\epsilon_{it}$ is a unit-level transitory shock. Similar models have been used to motivate synthetic control and diff-in-diff estimators (see \textcite{sdid}, \textcite{fermanpinto} or \textcite{abadievives}), as well as two-way fixed effect estimators and interactive fixed effect estimators (\textcite{bai2009}, \textcite{gobillon}). Assumption 1 imposes restrictions on the model primitives common in the literature.

\begin{assumption}[Model primitives] Assumptions on the covariates, the factor structure and the error components.
\begin{itemize}
    \item The covariates are fixed and bounded such that $\max_j \|\bm{Z}_j\| \leq \sqrt{k}$. 
    \item The common factor $\lambda_{t}$ are covariance stationary. 
    \item $\epsilon_{it}$ are mean independent of $\{ \bm Z, \bm \mu\}$.
\end{itemize}
\end{assumption}

To study under which conditions the \textit{sparse} synthetic control algorithm selects the most important predictors, I assume a sparse representation of the predictors.

\begin{assumption}[Sparse representation] For all $t$, $\bm\theta_t$ is partitioned conformably into $(\tilde{\bm\theta}_t, \mathbf{0})'$ where $\tilde{\bm\theta}_t$ is a ($k_1\times1$) vector of non-zero parameters and $\mathbf{0}$ is a $k_2\times 1$ vector such that $k = k_1 + k_2$. 
\end{assumption}

Under Assumption 2 we partition the covariates comformably into a non-zero and zero group such that $\bm Z_i = (\bm Z_i^1, \bm Z_i^2)$. Throughout the paper I refer to the $\bm Z_i^1$ predictors as the "useful" predictors and the $\bm Z_i^2$ predictors as the nuisance predictors. Intuitively, we say that the \textit{sparse} synthetic control is \textit{interpretable} if given a large set of predictors the algorithm successfully recovers the useful ones. That is, we would like to require the method to consistently select which covariates are useful and give them positive weight, while giving zero weight to the nuisance covariates. Without further assumptions, however, this requirement is too strong. Indeed, if the nuissance covariates are highly correlated with the useful covariates then there is little hope to achieve model consistency. Assumption 3 gives a strong condition to avoid this problem. 

\begin{assumption}[Oracle covariate match] For fixed $J$, let the oracle weights be defined by
$$
\bm{w}^* \in \text{argmin}_{w\in \Delta^J} \mathbb{E} \|\bm Y_1 - \bm Y_0\bm w\|^2.
$$ 
\noindent We consider two assumptions:
\begin{enumerate}
    \item For all $k\in S = \{k \mid \theta_{tk} = 0 \text{ for all } t\}$, $|Z_{1k} - Z_{Jk}'\bm{w}^*| > 0$.
    \item (1) holds true and for $l\in S^c$, $|Z_{1l} - Z_{Jl}'\bm{w}^*| =0$.
\end{enumerate} 
\end{assumption}

Assumption 3 is different to the irrepresentability condition and the restricted eigenvalue condition common in compressed sensing and model selection consistency for penalized estimators (see \textcite{zhaoyu} and \textcite{chernlasso} for applications to lasso estimators). It says that the weights that minimize the statistical risk can not lead to a perfect match for a nuisance covariate. This assumption rules out data generating processes for which matching nuisance covariates improves the overall fit. For example, this could happen if the nuisance covariates are correlated with the factor loadings and contain useful information for matching them. Without this assumption, there could be cases in which a nuisance covariate is actually useful in improving the outcome fit and also can be matched at no cost to the lower level program, making it desirable for our algorithm to use it with positive probability.

The main result, Theorem 1, states that the \textit{sparse} synthetic control algorithm is model consistent when the number of pre-treatment periods grows and $J$, $k$ and $F$ are fixed.

\begin{theorem}[Model Selection]
Under A1-A3.1 if $\psi$ is an injective function and $\hat{\lambda} \to 0$ as $T_0 \to \infty$, for a fixed $k$ and $J$, as $T_0 \to \infty$ the following holds\footnote{Here, $\hat{\lambda}$ is the cross-validated hyper-parameter of the penalty function denoted by $\lambda^*$ in Algorithm 1.}
\begin{enumerate}
    \item If $k\in S = \{k \mid \theta_{tk} = 0 \text{ for all } t\}$, then $P(v_k = 0) \to 1$.
    \item If A3.2 holds and $l \in S^c$ then $P(v_l = 0) \to 0$.
\end{enumerate}
\noindent where $v_k$ is the predictor weight for predictor $m$ assigned by the sparse synthetic control algorithm.
\end{theorem}

In words Theorem 1 states that the \textit{sparse} synthetic control algorithm will not use the nuisance predictors with probability going to one as the number of pre-treatment periods increases. The two main assumptions to derive the result are the oracle selection assumption and the assumption that the bi-level program has a unique optimum such that $\psi$ is injective. This will be the case when the data points are in general position, that is, when every subset of the columns of $\bm X$ are linearly independent. These two assumptions are hard to test in practice.

It is important to note that the implication does not go the other way in general. The method could set a useful predictor weight to zero, but it will not assign a nuisance predictor a non-negative weight with high probability. Only if the useful predictors can be matched perfectly in the population, then the method is sign consistent. This result is important because it justifies the use of \textit{sparse} synthetic controls to identify important predictors. A researcher that is unsure about what predictors to use to generate the synthetic control can use the \textit{sparse} synthetic control with many predictors and be confident that it will use the "useful" predictors if the oracle separation assumption is satisfied. This suggests using the method as an alternative to manual \textit{predictor search} by researchers and possibly as a way to prevent specification search.

Next, we show that the \textit{sparse} synthetic control also has desirable performance properties when compared to the standard (un-penalized) method. In particular, it has a faster convergence rate the for bias and mean-squared-error of the treatment effect estimator. To show this, recall that the treatment effect of interest is $\tau_{1t} = Y_{1t}^I - Y_{1t}^N$ for $t>T_0$. To estimate it we generate a counterfactual $\hat{Y}^N_{1t} = \sum_{j=2} ^{J+1}w_jY_{jt}^N$ for a synthetic control $\mathbf{w} = (w_2, \dots, w_{J+1})'$. Therefore, the estimated treatment effect indexed by a synthetic control $\bm w$ is given by

$$
\hat{\tau}^{\bm w}_{1t} = \bm{\theta}_t' \left( \bm Z_1 - \sum_{j=2} ^{J+1}w_j\bm Z_j\right) + \bm{\lambda}_t' \left( \bm{\mu}_1 - \sum_{j=2} ^{J+1}w_j\bm{\mu}_j\right) + \sum_{j=2} ^{J+1}w_j (\epsilon_{1t} - \epsilon_{jt}).
$$

It is well known that when the synthetic control does not have perfect pre-treatment fit it is biased. The following lemma gives an upper bound on the bias that depends on two terms that do not vanish asymptotically. 

\begin{lemma}[Bias Bound]
Under A1-A2 and assuming that there exists a $\bar{\lambda}$ such that  $|\lambda_{ft}| \leq \bar{\lambda}$ for all $t$ and $f$, and that the smallest eigenvalue of $\sum_{t=1}^{T_0} \bm{\lambda}_t'\bm{\lambda}_t$ is bounded below by $\underline{\xi}$. Then, for a synthetic control $\bm w$,

$$
\mathbb{E}|\hat{\tau}^{\bm w}_{1t}| \leq \frac{\gamma}{T_0} \sum_{m=1}^{T_0}\mathbb{E}|Y_{1m} - \sum_{j=2} ^{J+1}w_j Y_{jm}| + \left|\bar\theta\left( 1 - \frac{\gamma}{T_0} \right)\right| \sum_{k=1}^{k_1} \mathbb{E}|Z^1_{1k} - \sum_{j=2}^{J+1}w_jZ^1_{jk}| + \text{O}\left(T_0^{-1}\right).
$$

\noindent where $\gamma = \left(\frac{\bar{\lambda}^2 F}{\underline\xi}\right)$, $\bar \theta$ is the maximum value of $\tilde{\bm \theta}_t$, $F$ is the number of unobserved common factors and $k_1$ is the number of "useful" predictors. 
\end{lemma}

The bias bound in Lemma 1 provides three insights. First, if we don't have perfect pre-treatment fit the bias does not vanish asymptotically as $T_0$ increases. This means that our treatment effect estimate will be biased even if we have many pre-treatment periods. Second, the bias depends on the mean absolute deviation of the pre-treatment outcomes. Pre-treatment fit is, therefore, very important for controlling the bias. This leads to the suggestion of not using synthetic controls when the pre-treatment fit is bad. Third, and most relevant for the \textit{sparse} synthetic controls, the bias depends on the fit of the "useful" predictors $\bm Z_i^1$ and linearly in $k_1$. Therefore, even if we have a large number of pre-treatment periods and the pre-treatment fit is good (the first term in the bound is small), the bias could be large if the predictor fit is bad. Hence, a synthetic control that minimizes bias should attempt to perfectly match the useful predictors and disregard the nuisance predictors. This Lemma reinforces the result in \textcite{fermanpinto} that synthetic controls with imperfect pre-treatment fit may in general be biased.

To formalize the improvement in terms of bias and mean-square-error of the \textit{sparse} synthetic control over the standard synthetic control, we derive a finite sample rate for $\text{MSE}(\bm Z_0 \bm w)$ in the case in which the 

\begin{theorem}[MSE Rate]
Let $\bm Z_1 = \bm Z_0w^* + \bm u$ for $u_i \sim_{ind} \text{subG}(\sigma_z^2)$. Then, under $A1-A3.1$ as $T_0 \to \infty$, almost surely for the sparse synthetic control $\hat{\bm w}$, 
$$
MSE(\bm Z_1, \bm Z_0\hat{\bm w}) = \frac{1}{k} \|\bm Z_1- \bm Z_0\hat{\bm w}\|^2 \lesssim \frac{\sigma_z \sqrt{k_1}}{k} \sqrt{2 \log J}.
$$

\noindent For the standard synthetic control $\tilde{\bm w}$,
$$
MSE(\bm Z_1, \bm Z_0\tilde{\bm w}) = \frac{1}{k} \|\bm Z_1- \bm Z_0\tilde{\bm w}\|^2 \lesssim \sigma_z\sqrt{\frac{2 \log J}{k}}.
$$

\end{theorem}

Theorem 2 describes how the mean squared error of the predictor match changes with the number of predictors and donor units. The main difference between the \textit{sparse} synthetic control and the standard method is that the rate is of order $O(\sqrt{k_1}/k)$ instead of $O(1/\sqrt{k})$. This means that in a sparse setting, when $k_1$ is small with respect to $k$, the sparse synthetic control has a faster MSE rate. In practice, this is important because it implies that the sparse synthetic control will achieve lower MSE than the standard synthetic controls when both methods use the same number of predictors. More so, given Lemma 1, a faster predictor MSE rate also implies that the \textit{sparse} synthetic control will achieve lower bias than the standard method. Formally, recall that by Cauchy-Schwartz inequality the MAD is bounded above by $\sqrt{J}MSE$, so the proposed method achieves a lower MAD, and so bias, than the standard method. In section 4 we explore these properties further by estimating the expected mean absolute deviations empirically and show in a simulation study that the \textit{sparse} synthetic control achieves \textit{both} better outcome pre-treatment fit and better "useful" predictor fit than the standard synthetic control.

\section{Simulation Study}

In this section we study the performance of the \textit{sparse} synthetic control in relation to two benchmark models: the standard synthetic control with $\bm V$ chosen as $(\mathbf{X}_0^{train'}\mathbf{X}_0^{train})^{-1}$, which we label SCM, and the synthetic control with $\bm V$ chosen to minimize $\frac{1}{T_{val}} \|\mathbf{Y}_1^{val} - \mathbf{Y}_0^{val}\mathbf{w}(\mathbf{V})\|^2$ as proposed by \textcite{adh2015}, which we label SCM $\lambda=0$ as it can be understood as the unpenalized version of the \textit{sparse} synthetic control.

\noindent The findings below rely on the following simulation design, similar to the one used in \textcite{abadievives}, for $B=1000$ draws:\footnote{We also set $\delta_t = 100$, but without loss of generality $\delta_t$ could be set to zero.}
\begin{align*}
T &= 30, \text{ } T_0 = 20, \text{ } T_v = 10, \\
	\bm Z_i &= [\bm Z^1_{i}, \text{ } \bm Z^2_i], \text{ where } \bm Z^1_i, \bm Z^2_i \sim_{iid} U[0,1], \\
	\bm Z^1_1 &= \frac{1}{2} \bm Z^1_{2} + \frac{1}{2} \bm Z^1_{3}, \\
	\lambda_t &\text{ follows an } AR(1) \text{ with coefficient } \rho=0.5, \\
	\epsilon_{it} &\sim N(0, \sigma^2) \text{ with }\sigma =0.25, \\
	F&=7 \text{ in groups of 3 units and } J+1=21.
\end{align*}

This simulation design implies that unit 1 (the treated unit) can be perfectly replicated (up to noise) by an average of units 2 and 3. Hence, the optimal synthetic control would choose $w_2 = w_3 = \frac{1}{2}$. We study two different predictor settings. The first is one in which there are a similar amount of useful and useless predictors ($k_1 = k_2 = 5$); the second, includes only one "useful" predictor ($k_1=1$ and $k_2 = 9$). In both cases we add 10 lags of the outcome variable to the design matrix, for a total of 20 predictors. Note that this is a challenging design for the method as both useful and useless predictors are drawn from the same distribution. We summarize the simulation results in two Figures that show the performance of the different methods and explore the theoretical insights from Section 3.

\begin{figure}[H]
  \vspace{-3cm}
  \begin{subfigure}[b]{0.5\textwidth}
  	\hspace*{-1cm}
  	\includegraphics[width=1.2\linewidth]{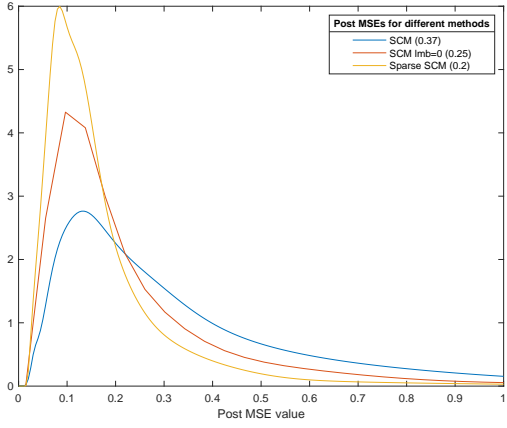}
  	\vspace*{-3.5cm}
  	\caption*{(a) $k_1 = k_2 = 5.$}
  \end{subfigure}
  \begin{subfigure}[b]{0.5\textwidth}
  	\hspace*{-1cm}
  	\includegraphics[width=1.2\linewidth]{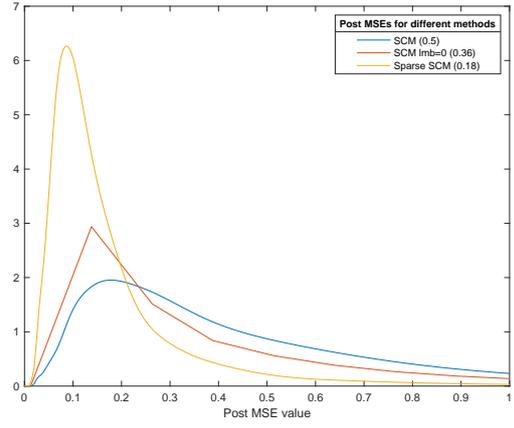}
  	\vspace*{-3.5cm}
  	\caption*{(b) $k_1 =1$, $k_2 = 9.$}
  \end{subfigure}
  \hspace*{-1cm}
   \begin{subfigure}[b]{0.5\textwidth}
     \vspace{-2.5cm}
  	\includegraphics[width=1.2\linewidth]{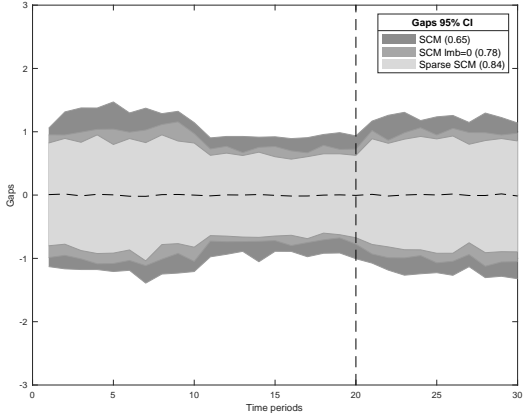}
  	\vspace*{-3.5cm}
  	\caption*{ \hspace*{1.5cm} (c) $k_1 = k_2 = 5.$}
  \end{subfigure}
  \begin{subfigure}[b]{0.5\textwidth}
    \vspace{-2.5cm}
  	\includegraphics[width=1.2\linewidth]{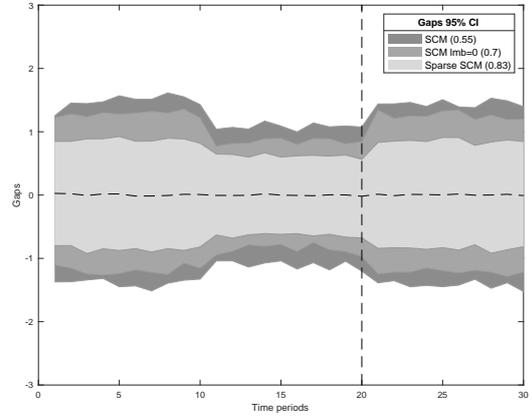}
  	\vspace*{-3.5cm}
  	\caption*{ \hspace*{1.5cm} (d) $k_1 =1$, $k_2 = 9.$}
  \end{subfigure}
  \caption{Mean Squared Errors.}
  \label{fig:mses}
    \begin{tablenotes}
		\small \item \textbf{Notes}: Panels (a) and (b) show the kernel density across simulations of the MSEs for the outcome variable in the post-treatment period, with average values in parenthesis. Panels (c) and (d) show the inter-quartile range of the $Y_{1t} - \hat{Y}_{1t}$ with $w^*_2 + w^*_3$ in parenthesis. SCM lmb=0 refers to the unpenalized synthetic control with $\lambda = 0$.
   \end{tablenotes}
\end{figure}

First, focus on the post-treatment mean squared error (MSE) of the outcome variable. Given that the MSE is an informative measure of fit that includes both bias and variance it gives us an idea of the performance of the estimator. In particular, lower post-treatment MSEs will imply lower standard errors for the treatment effect of interest. Figure \ref{fig:mses} shows the distribution of MSEs for the simulations for the two settings. Panels (a) and (b) show that the \textit{sparse} synthetic control has on average lower MSE and less dispersion than the two benchmark methods. Panels (c) and (d) show the distribution of the gaps between the predicted outcome and the real outcome for all periods. The \textit{sparse} synthetic control has tighter confidence intervals both before and after treatment and sets unit weights closer to the optimal control (83\% of the weight is given to units 2 and 3). Furthermore, observe that whereas the benchmarks methods perform worse in the $k_1=1, k_2=9$ setting, the \textit{sparse} synthetic control is able to perform similarly in both settings. This shows the ability of the method to perform well regardless of the number of useful and nuissance predictors. 

Recall that in section 3 the bias bound crucially depends on the $\text{MAD}$ of the pre-treatment outcomes and useful predictors. A figure in the appendix shows the $\mathbb{E} \text{MAD}$s for the outcome and useful predictors in the pre-treatment period. As in Figure 1, the \textit{sparse} synthetic control is able to perform well in both settings whereas the benchmark methods perform poorly in the $k_1=1, k_2=9$ setting.

\begin{figure}[H]
  \vspace{-3cm}
   \begin{subfigure}[b]{0.5\textwidth}
    \vspace{-2.5cm}
    \hspace*{-1cm}
  	\includegraphics[width=1.2\linewidth]{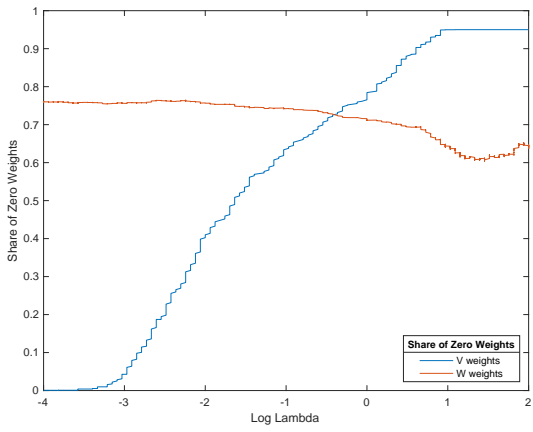}
  	\vspace*{-3.5cm}
  	\caption*{(c) $k_1 = k_2 = 5.$}
  \end{subfigure}
  \begin{subfigure}[b]{0.5\textwidth}
    \vspace{-2.5cm}
  	\hspace*{-1cm}
  	\includegraphics[width=1.2\linewidth]{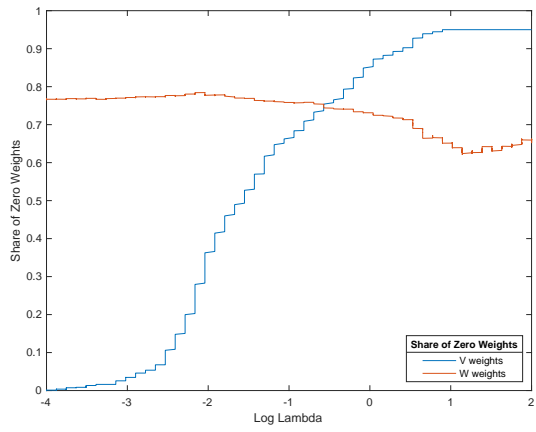}
  	\vspace*{-3.5cm}
  	\caption*{(d) $k_1 =1$, $k_2 = 9.$}
  \end{subfigure}
  \hspace*{-1.1cm}
  \begin{subfigure}[b]{0.5\textwidth}
     \vspace{-2.5cm}
  	\includegraphics[width=1.2\linewidth]{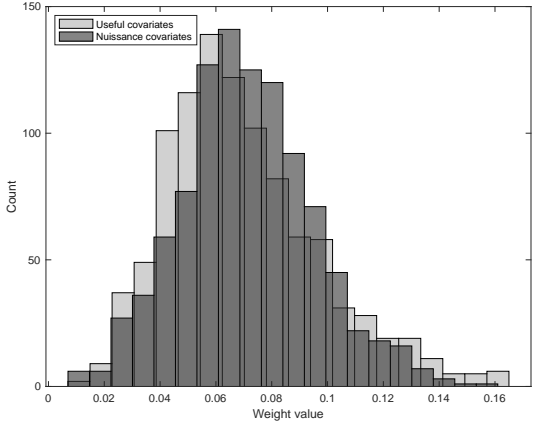}
  	\vspace*{-3.5cm}
  	\caption*{ \hspace*{1.5cm} (e) SCM $\lambda^* = 0$ $\bm V^*$}
  \end{subfigure}
  \begin{subfigure}[b]{0.5\textwidth}
    \vspace{-2.5cm}
  	\includegraphics[width=1.2\linewidth]{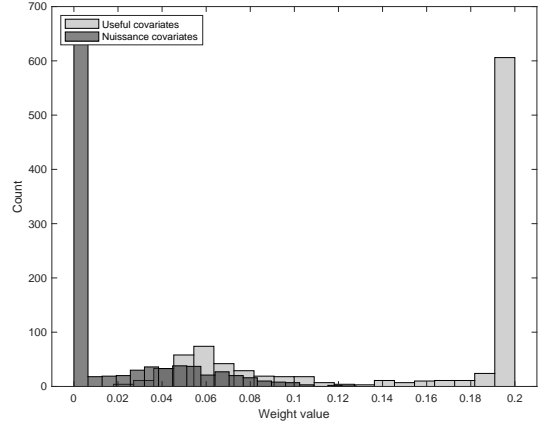}
  	\vspace*{-3.5cm}
  	\caption*{ \hspace*{1.5cm} (f) \textit{Sparse} SCM $\bm V^*$}
  \end{subfigure}
  \caption{Pre-treatment Fit and Variable Selection.}
  \label{fig:selection}
  \begin{tablenotes}
		\small \item \textbf{Notes}: Panels (a) - (b) show the share of the $\bm V$ and $\bm w$ weights that are zero for the different values of $\lambda^*$ across simulations for the \textit{sparse} synthetic control. Panels (c) - (d) show the histogram of the $v_k$ weights across simulations for the $k_1=k_2=5$ setting for the unpenalized synthetic control with $\lambda = 0$ and the \textit{sparse} synthetic control respectively.
   \end{tablenotes}
\end{figure}

To investigate the model selection result in Theorem 2 we plot the histogram of the predictor weight values ($v_k$) for the useful and useless predictors across simulations. Panels (c) and (d) in Figure \ref{fig:selection} compare the SCM with $\lambda^* =0$ and \textit{sparse} synthetic control for the $k_1=k_2=5$ setting. Whereas the standard SCM does not clearly distinguish between the useful and nuissance covariates, the \textit{sparse} synthetic control correctly assigns zero weight to the nuissance covariates most of the time. The stark difference between the two models suggests that the \textit{sparse} synthetic control may be able to perform variable selection in practice. Finally, in panels (a) and (b) in Figure \ref{fig:selection} we show that the \textit{penalized} synthetic control induces sparsity when the $\lambda$ hyper-parameter increases, but that the donor weights ($w_j$) remain stable. The main takeaway is that the unit weights $\bm w^*$ have the same amount of sparsity regardless of the magnitude of the optimal regularization. This is evidence to motivate the technical assumption that $\psi$ is injective in Theorem 2, and confirms that the method can reliably achieve a unique optimum. Note that we do see some instability for large values of $\lambda^*$ in cases in which only one predictor is used.

\section{Extending the California smoking program case study}

\begin{figure}[H]
  \vspace{-3cm}
  \hspace*{-2cm}
  \begin{subfigure}[b]{0.5\textwidth}
  	\includegraphics[width=1.35\linewidth]{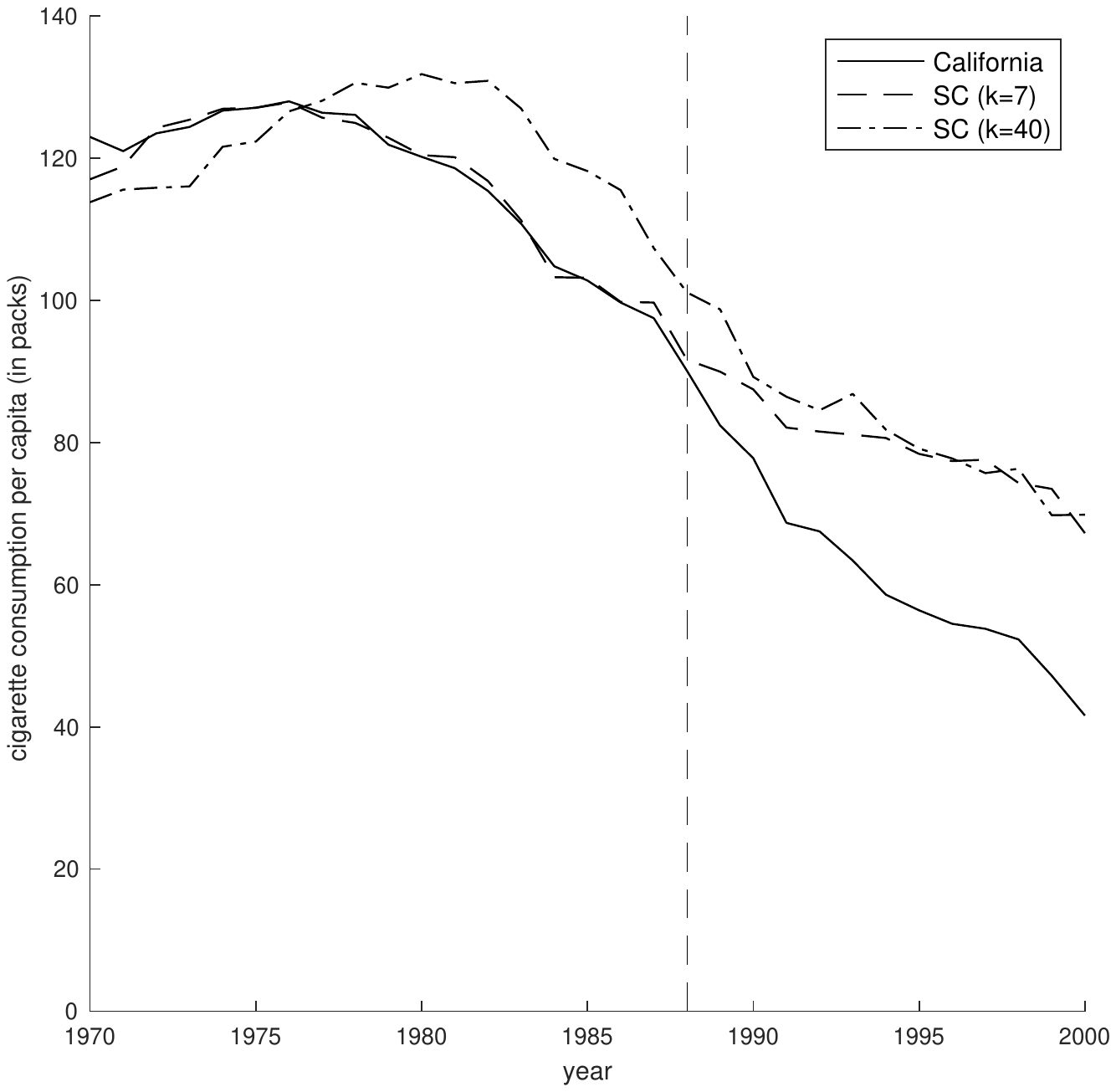}
  	\vspace*{-3.5cm}
  	\caption*{ \hspace*{2cm} (a) Standard SC}
  \end{subfigure}
  \begin{subfigure}[b]{0.5\textwidth}
  	\includegraphics[width=1.35\linewidth]{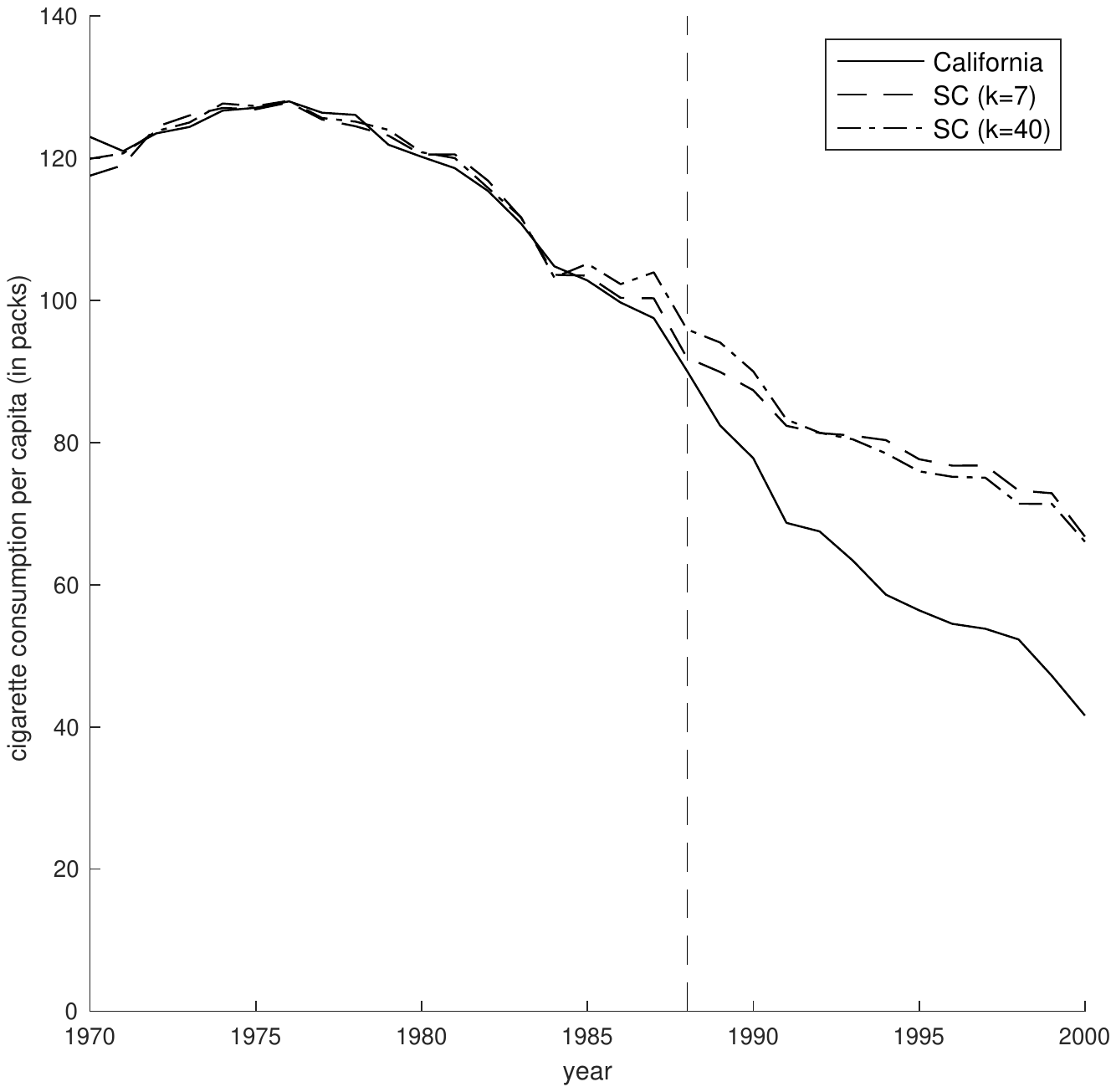}
  	\vspace*{-3.5cm}
  	\caption*{ \hspace*{2cm} (b) \textit{Sparse} SC}
  \end{subfigure}
  \caption{Synthetic Controls for California.}
  \label{fig:cali_sc}
\end{figure}

In 1988 proposition 99 increased California’s cigarette excise tax by 25 cents per pack and shifted public policy towards a clean air agenda. This policy intervention has been used extensively to compare the validity and performance of various synthetic controls and diff-in-diff estimators. The outcome of interest is cigarette sales per capita in packs in California and the donor pool includes 38 states without similar policy interventions. The original data set used in the \textcite{adh2010} study used seven predictors: Ln(GDP per capita), percent aged 15–24, retail price, beer consumption per capita and three lags of cigarette sales per capita (1975, 1980 and 1988). In the study the standard synthetic control built using this design matrix falls outside the convex hull of the predictors, but has very good pre-treatment fit. This can be seen in Figure \ref{fig:cali_sc} panel (a) for the standard synthetic control with $k=7$.

To study the potential benefits of using penalized synthetic control methods, we augment the original predictor data set with 33 additional predictors. The additional predictors are obtained from the IPPSR (MSU) dataset assembled by \textcite{msu} of policy correlates in the United States, they include demographic variables, income related variables, political participation measures and government spending statistics. Importantly, they also include less useful variables such as state identifiers, but do not include predictors that do not vary across states as these would violate A3.1. Figure \ref{fig:cali_sc} compares the standard synthetic control (SCM with $\lambda^*=0$) and the synthetic control calculated using Algorithm 1 using the $k=7$ predictors considered in \textcite{adh2010} and the extended set of predictors ($k=40$).\footnote{To choose the optimal predictor weights, we divide the pre-treatment period in a training period (14 years) and a validation period (5 years).}

Figure \ref{fig:cali_sc} highlights the difficulty of the standard synthetic control matching problem when the number of predictors is large. While a good synthetic control exists in the $k=40$ case, namely the same as the synthetic control for $k=7$, the standard procedure may be unable to find a global optima and fail to recover the set of predictors for which a good match that concurrently optimizes pre-treatment fit exists. The synthetic control reported in Figure \ref{fig:cali_sc} panel (a) for $k=40$ does not have good pre-treatment fit, and therefore, may lead to biased estimates of the average treatment effect on the treated. This example reinforces the notion in practice the choice of predictor set is very important in constructing valid synthetic controls. If researchers do not want to manually choose the predictors to match, or may not have a good prior on which predictors might be useful, the \textit{sparse} synthetic control method we propose may be an attractive alternative. As can be seen in panel (b) of Figure \ref{fig:cali_sc} the sparse procedure is able to find a good synthetic control in the original and extended settings. 

The theoretical results and simulation evidence suggest that the \textit{sparse} synthetic control procedure should perform better in terms of bias and prediction mean squared error than the standard method in settings in which the predictors have a sparse representation. To explore this in the context of the California case study, Table \ref{tab:cali} reports the average treatment effect on California of the passage of proposition 99 over the 11 post-treatment years and its placebo variance estimate for the Differences-in-Differences estimator using the rest of the US as control group, the standard synthetic control (SCM with $\lambda^*=0$), and the \textit{sparse} synthetic control for the original and extended predictor settings. The placebo variance is estimated using the placebo bootstrap method proposed in \textcite{sdid}, a detailed explanation of the method can be found in the appendix. 

\begin{table*}
	\caption{Treatment effect estimates.}
	\begin{center}
 	\begin{tabular}{||c |c | c| c| c| c ||} 
 	\hline
 	& DID & SCM & \textit{Sparse} SCM & SCM  & \textit{Sparse} SCM \\ [0.5ex] 
 	\hline\hline
 	$\hat{\tau}$ estimate & -27.4 & -18.9 & -18.5 & -21.0 &  -18.2 \\ 
 	$\hat{V}^{1/2}_{\tau}$ & (16.7) & (13.2) & ( 12.2 ) & ( 12.9 ) &  ( 11.7 ) \\
        k & - & 7 & 7 & 40 & 40 \\
 	\hline
	\end{tabular}
	\end{center}
        \label{tab:cali}
	\begin{tablenotes}
		\small \item \textbf{Notes}: DID is the standard diff-in-diff estimator, SCM is the standard synthetic control with $\bm V$ chosen to minimize $\frac{1}{T_{val}} \|\mathbf{Y}_1^{val} - \mathbf{Y}_0^{val}\mathbf{w}(\mathbf{V})\|^2$ without penalization ($\lambda^* = 0$), SDID is the synthetic diff-in-diff estimator, and the $'+'$ indicates the augmented data setting. Standard errors are taken from Arkhangelsky et al. 2020.		\end{tablenotes}
\end{table*}

To benchmark the findings in Table \ref{tab:cali} in the first column we report the Differences-in-Differences estimate which, given that the parallel trends assumption is likely violated, is biased downwards as discussed in \textcite{adh2010}. The second column shows the standard synthetic control with the seven predictors as in \textcite{adh2010}, which given its excellent pre-treatment fit, is believed to have low bias. Column 3 shows that, unsurprisingly, the \textit{sparse} method yields a very similar treatment effect as the standard method when $k=7$. The difference between the methods become apparent in the last two columns. While the standard method becomes biased towards the differences-in-differences estimate when $k=40$, the penalized method obtains a very similar estimate as in the $k=7$ case. This robustness to the number of predictors highlights the benefits of the penalized procedures in choosing which predictors to match on to optimize the pre-treatment fit while avoiding over-fitting. Furthermore, we also see an improvement in the placebo variance estimates between the standard method and the penalized method, with the later having 8\%-10\% lower variance. 

\begin{figure}
\caption{Predictor choice in the $k=40$ California case study.} 
\label{fig:pred_cal}
\begin{minipage}{0.5\textwidth}
\subcaption{Top 7 predictors}
\begin{tabular}{ll}
\toprule
                      SCM &       \textit{Sparse} SCM \\
\midrule
                  smk\_80 &          smk\_75 \\
     general\_revenue\_inc &   incshare\_top1 \\
                  smk\_75 &          smk\_88 \\
                  smk\_88 &      pc\_inc\_ann \\
                  loginc &          region \\
 general\_expenditure\_inc &  budget\_surplus \\
              pc\_inc\_ann &       taxes\_gsp \\
\bottomrule
\end{tabular}
\end{minipage}
\hfill
\begin{minipage}{0.45\textwidth}
    \subcaption{Predictor weight distribution}
    \label{fig:EWEBdis}
    \includegraphics[width=1.3\textwidth]{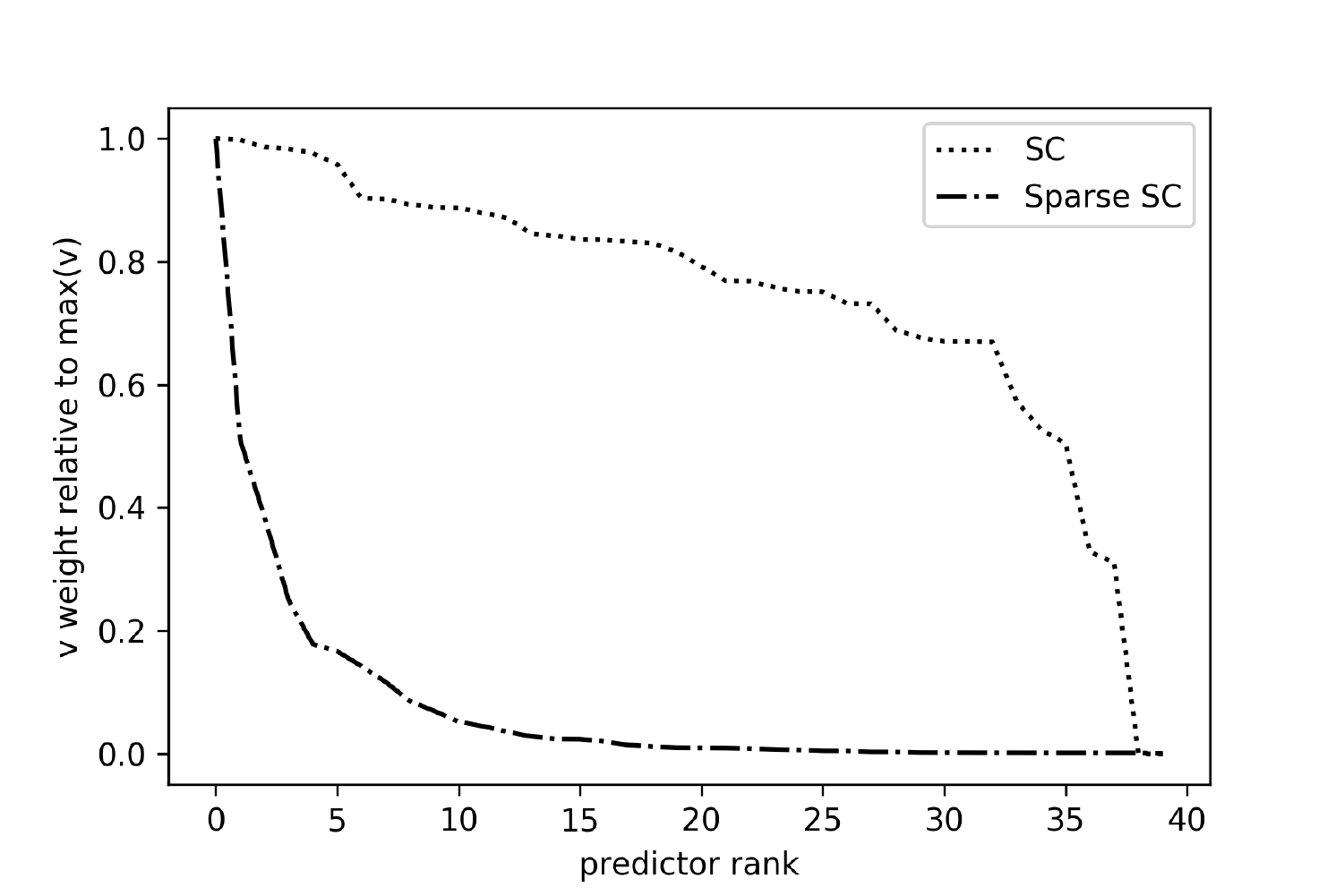}
\end{minipage}
\end{figure}

Finally, in Figure \ref{fig:pred_cal} we explore the predictor weight choices of each method when $k=40$. Panel (b) shows that, as expected, the penalized method uses predominantly a small set of predictors, with a large fraction of predictors receiving zero or close to zero weight. This comes in stark opposition to the standard method that uses denser predictor weights, with many predictors having similar weight and only a few being set to zero. Panel (a) confirms that the choice of predictors in \textcite{adh2010} was very good. Amongst the top seven predictors with highest weight, both methods use the lagged cigarette sales variables (smk) and personal income related variables log income and per capita annual income. Departing from the \textcite{adh2010} predictor set both methods also assign weight to state level revenue and spending variables such as the general revenue, the budget surplus or the tax revenue as a percentage of the gross state product and both methods assign close to zero weight to the beer consumption measure. Additionally, the sparse method also uses a regional indicator (South-West-Midwest-East) and an inequality measure (the share of income by the top 1\%). Finally, the sparse method assigns zero weight to the state indicator (a predictor we expected not to be useful in the match), while the standard method uses the indicator with positive weight. Suggesting that the model selection result for the sparse method is relevant in applied settings in identifying nuisance predictors.

The main takeaways from the empirical application are (1) that the standard SCM may struggle to generate a valid synthetic control, and so may be biased, when we use a large number of predictors, (2) that the \textit{sparse} synthetic control is a potential solution to this problem and is robust to increasing the number of predictors, and (3) that using the penalized method in combination with a larger set of predictors may be useful for researchers that have many predictors at their disposal and are unsure about which ones may be the useful ones.

\section{Conclusion}

Researchers and policy makers are increasingly drawn towards synthetic control methods to analyze policy interventions. A key step in using these methods is deciding what predictors to use in building the synthetic control. Using a large number of predictors is not a valid alternative as it can lead to biased treatment effects and poor post-treatment performance. In this paper, we advocate for a data-driven penalized synthetic control method that automatically chooses the important predictors. We suggest that this method can be used as an alternative to the standard synthetic control when the researcher has many predictors at his disposal but does not know which ones should be used.

We show that, in a linear factor model setting, that the \textit{sparse} synthetic control is model consistent and can successfully recover which predictors are useless when the number of pre-treatment periods is large. Furthermore, by deriving an MSE convergence rate result and a bias bound result, we show that the \textit{sparse} synthetic control has better theoretical performance properties than the un-penalized method. Motivated by this insight we then show through a simulation study that the proposed method is able to reduce both the bias and the mean-squared error measures with respect to the un-penalized synthetic control. 

Finally, we highlight the practical relevance of the method by applying it to the passage of Proposition 99 in California in a setting with 8 predictors versus a setting with 40 predictors. Whereas the standard synthetic control estimate becomes more biased when the number of predictors is increased, the \textit{sparse} synthetic control is robust to the number of predictors used. Through this exercise, we confirm that the choice of predictors in the original study of \textcite{adh2010} was good, as the penalized method uses the original predictors with positive probability.

A natural next step for future work is to explore alternative penalized synthetic control methods with multidimensional shrinkage (\textcite{vives-i-bastida_2021}), for example with elastic net penalties, that may further improve the performance of the method and the standard errors of the treatment effect estimates.

\newpage
\printbibliography
\newpage

\section*{Appendix}

\subsection{Note on computational problem}

Consider the lower level problem:
$$
 \psi(\mathbf{V}) = \text{argmin}_{\mathbf{w}\in \mathcal{W}} L_W(\mathbf{V},\mathbf{w}) = \|\mathbf{X}^{train}_1 - \mathbf{X}^{train}_0\mathbf{w}\|^2_V.
$$
The objective function can be re-written as follows, dropping the \textit{train} label for ease of notation
\begin{align*}
    \|\mathbf{X}_1 - \mathbf{X}_0\mathbf{w}\|^2_V &= (\mathbf{X}_1 - \mathbf{X}_0\mathbf{w})'V(\mathbf{X}_1 - \mathbf{X}_0\mathbf{w}) \\
    &= \sum_k x_{1k}^2 v_k - 2 \sum_j w_j(\sum_k x_{jk}v_kx_{1k}) + \sum_i w_i \sum_k x_{ik}v_k(\sum_j x_{jk}w_j).
\end{align*}

Then, the derivative with respect to $w_j$ is given by
$$
\frac{\partial L_W}{\partial w_j} = -2(\sum_k x_{jk}v_kx_{1k}) + \sum_k x_{jk}v_k(\sum_i x_{ik}w_i) + \sum_i w_i \sum_k x_{ik}v_kx_{jk}
$$
and the derivative with respect to $v_k$
$$
\frac{\partial L_W}{\partial v_k} = x_{1k}^2 - 2\sum_j w_jx_{jk}x_{1k} + \sum_i w_i x_{ik}(\sum_j x_{jk}w_j).
$$

Oberve that this implies that while the magnitude of $\psi(\mathbf{V})$ changes with the scale of $\mathbf{V}$ the minimizers $\mathbf{w}$ satisfying $\frac{\partial \psi(\mathbf{V})}{\partial w_j} = 0$ do not. Indeed, multiplying all $v_k$s by a scalar $a>0$ will not change the set of minimizers. This implies that we can not jointly minimize the upper and lower level problems directly with the penalty $\lambda \| \text{diag}(\mathbf{V})\|_1$, as $\| \text{diag}(\mathbf{V})\|_1 \to 0$ is optimal when $v_k\geq 0$. This is why in Algorithm 1 we set a $v_{k} = 1$ before the minimization step.

If $v_{k_0} = 1$ for some $k_0 \in \{1, \dots, K\}$, then 
\begin{align*}
    \frac{\partial L_W}{\partial w_j} = -2(x_{jk_0}x_{1k_0} &+ \sum_{k\neq k_0} x_{jk}v_kx_{1k}) + x_{jk_0}(\sum_i x_{ik}w_i) +\sum_{k\neq k_0} x_{jk}v_k(\sum_i x_{ik}w_i) \\
    &+ \sum_i w_i \sum_{k\neq k_0} x_{ik}v_kx_{jk} + \sum_i w_i x_{ik_0}x_{jk_0}.
\end{align*}

In this case, the scale of $\mathbf{V}$ is relative to the $\mathbf{X}_{k_0}$, so re-scaling the $v_k$ for $k\neq k_0$ by a positive constant will change the set of minimizers given by $\{\frac{\partial L_W}{\partial w_j} =0\}$. In practice, the choice of $k_0$ is meaningful as every $k_0$ will yield slightly different $\mathbf{w}_j^*$ and the predictor $k_0$ will be included with probability one in the model. Researchers may choose a $k_0$ that they know to be important (i.e. a covariate that should be included in the model given domain knowledge), or may treat the $k_0$ as a hyper-parameter and iterate over all possible $k$s and choose the $k_0$ that minimizes the upper level loss.

\subsection{Proof of Lemma 1}

Consider the linear factor model described in section 3. For notational convenience in this section we do not use boldface terms, and following \textcite{adh2010} we use $Y^P$ to denote the vector of outcomes for the pre-treatment period. Our counterfactual will be a weighted average of the outcome variable for the donor pool

$$
\sum_{j=2} ^{J+1}w_jY^N_{jt} = \delta_t + \theta_t \sum_{j=2} ^{J+1}w_jZ_j + \lambda_t \sum_{j=2} ^{J+1}w_j\mu_j + \sum_{j=2} ^{J+1}w_j \epsilon_{jt}.
$$

As a result, the treatment effect $\tau_{1t}^w = Y_{1t} - \sum_{j=2}^{J+1}w_jY^N_{jt}$ for some weight vector $w = \{w_j\}_{j=2}^{J+1}$ takes the form
$$
\tau^w_{1t} = \theta_t \left( Z_1 - \sum_{j=2} ^{J+1}w_jZ_j\right) + \lambda_t \left( \mu_1 - \sum_{j=2} ^{J+1}w_j\mu_j\right) + \sum_{j=2} ^{J+1}w_j (\epsilon_{1t} - \epsilon_{jt}).
$$

\noindent Under some additional definitions and mean independence of the error term we have that:

\begin{align*}
	\tau_{1t}^w &= \lambda_t (\lambda^{P'}\lambda^P)^{-1}\lambda^{P'}\left(Y_1^P - \sum_{j=2} ^{J+1}w_j Y_j^P\right) \\
	&+ \left(\theta_t - \lambda_t(\lambda^{P'}\lambda^P)^{-1}\lambda^{P'}\theta^P\right)\left(Z_1 - \sum_{j=2} ^{J+1}w_j Z_j\right) \\
	& - \lambda_t (\lambda^{P'}\lambda^P)^{-1}\lambda^{P'}\left(\epsilon_1^P - \sum_{j=2} ^{J+1}w_j \epsilon_j^P\right) \\
	& + \sum_{j=2} ^{J+1}w_j (\epsilon_{1t} - \epsilon_{jt}).
\end{align*}

\noindent In \textcite{adh2010}, a bias bound is derived in the case in which we perfectly replicate the treated unit and the first two terms are zero. They show that the last two terms are $\text{O}\left(\frac{1}{T_0}\right)$. This paper explores the case in which we have many covariates and therefore are likely to fall outside the convex hull of the donor pool. In such cases, the synthetic control will not be able to replicate the design matrix of the treated unit and the pre-treatment fit will not be perfect. 

Assume that $\sum_{t=1}^{T_0} \lambda_t'\lambda_t$ is positive semi-definite with smallest eigenvalue bounded away from zero by $\underline{\xi}$ and $|\lambda_{tf}|< \bar{\lambda}$,  $|\theta_{tf}|< \bar{\theta}$ for all $t,f$. Then, it can be shown by the C-S inequality that:
$$
\left(\lambda_t \left(\sum_{s=1}^{T_0} \lambda_t' \lambda_t\right)^{-1} \lambda_m'\right)^2 \leq \left(\frac{\bar{\lambda}^2 F}{T_0 \underline\xi}\right)^2.
$$

\noindent Then, consider the first term in the decomposition of the treatment effect,

\begin{align*}
	\lambda_t (\lambda^{P'}\lambda^P)^{-1}\lambda^{P'}\left(Y_1^P - \sum_{j=2} ^{J+1}w_j Y_j^P\right)  &= \sum_{m=1}^{T_0} \lambda_t \left( \sum_{s=1}^{T_0} \lambda_t' \lambda_t\right)^{-1}\lambda_m'(Y_{1m} - \sum_{j=2} ^{J+1}w_j Y_{jm}) \\
	&\leq \sum_{m=1}^{T_0} \left| \lambda_t \left( \sum_{s=1}^{T_0} \lambda_t' \lambda_t\right)^{-1}\lambda_m' \right| \left|Y_{1m} - \sum_{j=2} ^{J+1}w_j Y_{jm}\right|\\
	&\leq \left(\frac{\bar{\lambda}^2 F}{T_0 \underline\xi}\right) \sum_{m=1}^{T_0}\left|Y_{1m} - \sum_{j=2} ^{J+1}w_j Y_{jm}\right| \\
	&= \left(\frac{\bar{\lambda}^2 F}{\underline\xi}\right)  \text{MAD}\left(Y_1^P,\sum_{j=2} ^{J+1}w_j Y_j^P\right).
\end{align*}

\noindent Therefore, the bias contribution from the first term in the $\tau_{1t}^w$ decomposition (which we denote by $R_{1t}$) is bounded by:
$$
\mathbb{E}|R_{1t}| \leq \left(\frac{\bar{\lambda}^2 F}{\underline\xi}\right)  \mathbb{E}\text{MAD}\left(Y_1^P,\sum_{j=2} ^{J+1}w_j Y_j^P\right).
$$

Similar, the second term is given by

 \begin{align*}
	\left(\theta_t - \lambda_t(\lambda^{P'}\lambda^P)^{-1}\lambda^{P'}\theta^P\right)\left(Z_1 - \sum_{j=2} ^{J+1}w_j Z_j\right) &= \sum_{k=1}^K \left(\theta_{tk} - \sum_{l=1}^{T_0} \lambda_t \left(\sum_{s=1}^{T_0} \lambda_s'\lambda_s\right)^{-1}\lambda_l'\theta_{lk}\right)\left(Z_{1k} - \sum_{j=2}^{J+1}w_jZ_{jk}\right) \\
	&\leq \sum_{k=1}^K \left|\theta_{tk} - \sum_{l=1}^{T_0} \lambda_t \left(\sum_{s=1}^{T_0} \lambda_s'\lambda_s\right)^{-1}\lambda_l'\theta_{lk}\right|\left|Z_{1k} - \sum_{j=2}^{J+1}w_jZ_{jk}\right| \\
	&\leq \left|\bar\theta\left( 1 - \left(\frac{\bar{\lambda}^2 F}{T_0 \underline\xi}\right)\right)\right| \sum_{k=1}^K \left|Z_{1k} - \sum_{j=2}^{J+1}w_jZ_{jk}\right| \\
	&= K \left|\bar\theta\left( 1 - \left(\frac{\bar{\lambda}^2 F}{T_0 \underline\xi}\right)\right)\right|  \text{MAD}\left(Z_1,\sum_{j=2} ^{J+1}w_j Z_j\right).
\end{align*}

The bias contribution from the second term ($R_{2t}$) is then given by:
$$
\mathbb{E}|R_{2t}| \leq K \left|\bar\theta\left( 1 - \left(\frac{\bar{\lambda}^2 F}{T_0 \underline\xi}\right)\right)\right|  \mathbb{E}\text{MAD}\left(Z_1,\sum_{j=2} ^{J+1}w_j Z_j\right).
$$

Given that the other two terms are $\text{O}\left(\frac{1}{T_0}\right)$ the result follows.

\subsection{Proof of Theorem 1}
Start by re-writing the bi-level optimization problem as a convex optimization problem where the lower level problem solutions are characterized by the necessary Karush-Kuhn-Tucker (KKT) conditions. The program can therefore be re-written as

\begin{align*}
    \min_{\bm w, \bm v, \kappa^w_1, \boldsymbol{\kappa}^w_2}\quad  &\frac{1}{T_{val}} \|\bm Y_1^{val} - \bm Y_0^{val}\bm w\|^2 + \lambda(1+\sum_{k\neq k_0}^K v_{k})\\
    s.t.\quad &  2\mathbf{X}_0^{train'}\mathbf{V}\mathbf{X}_0^{train}\mathbf{w} -2\mathbf{X}_0^{train'}\mathbf{V}\mathbf{X}_1^{train} + \kappa_{1}^w\mathbf{1}_J + \boldsymbol{\kappa}^w_2 = \mathbf{0}, \\
    & \sum_j \kappa^w_{2j}w_j = 0, \\
    & \kappa^w_{2j} \geq 0\text{ for all $j$}, \\
    & v_k\geq 0 \text{ for all $k\neq k_0$}
\end{align*}
\noindent where $\kappa_{1}^w$ is the Lagrangian multiplier for the sum to 1 constraint and $\bm \kappa_{2}^w$ are the associated multipliers for the non-negative constraint for the $w_j$ terms in the inner problem.

\noindent The associated Lagrangian for parameter vector $\bm \nu =( \bm w, \bm v, \kappa^w_1, \boldsymbol{\kappa}^w_2, \boldsymbol{\kappa}^v, \bm \xi_1, \xi_2, \bm\xi_3)'$ is 

\begin{align*}
    \min_{\bm \nu} \mathcal{L}(\bm \nu) \equiv& \quad \frac{1}{T_{val}}\|\bm Y_1^{val} - \bm Y_0^{val}\bm w\|^2 + \lambda(1+\sum_{k\neq k_0}^K v_{k})\\
    \quad & +\bm\xi'_1(2\mathbf{X}_0^{train'}\mathbf{V}\mathbf{X}_0^{train}\mathbf{w} -2\mathbf{X}_0^{train'}\mathbf{V}\mathbf{X}_1^{train} + \kappa_{1}^w\mathbf{1}_J + \boldsymbol{\kappa}^w_2) \\
    &+ \xi_2\sum_j \kappa^w_{2j}w_j \\
    &+ \xi_3\boldsymbol{\kappa}^w_{2} \\
    &+ \bm\kappa^{v'} \bm v.
\end{align*}

\noindent subject to the KKT dual feasibility and slackness conditions:
\begin{align*}
    \text{\textbf{Complementary slackness}}:\quad &\sum_{k\neq k_0} \kappa^v_{k}v_k = 0,\\
    &\sum_j \xi_{3j}\kappa_{j}^w =0.\\
    \text{\textbf{Dual feasibility}}:\quad &\quad \bm \kappa^v \geq \bm0,\quad \bm \xi_{3}\geq \bm 0,
\end{align*}
\noindent where $\kappa^v_1$ $\bm \kappa^v_{2}$ are the associated Lagrangian multipliers for the constraints on $v_k$ and $\bm \xi_1, \xi_2$,  and $\bm\xi_3$ are the multipliers associated with the lower level KKT necessary conditions. 

\noindent Observe that the lower level loss first order condition is given by:
$$
\mathbf{X}_0^{train'}\mathbf{V}\mathbf{X}_0^{train}\mathbf{w} -\mathbf{X}_0^{train'}\mathbf{V}\mathbf{X}_1^{train} = \sum_k \mathbf{x}_k \mathbf{x}_k'v_k\cdot \mathbf{w} - \sum_k \mathbf{x}_k x_{1k}v_k.
$$
\noindent Taking the derivative with respect to $v_k$ yields
$$
\frac{\partial}{\partial v_k} \implies \mathbf{x}_k( \mathbf{x}_k'\mathbf{w} - x_{1k}).
$$

\noindent It follows that the derivative of the objective function of $\mathcal{L}(\bm \nu)$ with respect to $v_k$ is given by
\begin{align*}
    \frac{\partial \mathcal{L}(\bm \nu)}{\partial v_k} &= \xi_{1}'\mathbf{x}_k( \mathbf{x}_k'\mathbf{w} - x_{1k}) + \kappa_{k}^v + \lambda.
\end{align*}

Under the assumption that that $\psi: \mathcal{V} \to \mathcal{W}$ is an injective function, it follows that there is a unique solution to the lower level optimization problem. This is the case when every subset of $k$ columns of $\bm X_0$ spans $\mathbb{R}^k$ (that is, the columns in every subset are linearly independent) and when $\bm V$ is diagonal and positive definite so that $\bm X_0'\bm V \bm X_0$ is non-singular. Since $\lambda, \kappa_k^v \geq 0$, the FOC  implies that $\rho_k \equiv  \xi_{1}'\mathbf{x}_k( \mathbf{x}_k'\mathbf{w} - x_{1k}) \leq 0$. Given a choice of $\lambda$, by the complementary slackness condition $\kappa_v^k>0$ for all $k$ such that $\rho_k  + \lambda = 0$ and $\kappa_v^k=0$ otherwise. Therefore, the choice of $\lambda$, together with $\bm w$ and the multipliers $\xi_1$, control how many predictors will receive zero weight.

We will now show that as $T_0, T_{val} \to \infty$, the $\rho_k$ for $k \in S = \{k \mid \theta_{kt} = 0 \text{ for all } k,t\}$, do not converge to zero in probability. Consider using the predictors $Z$ for the design matrix $\bm X$, such that $\rho_k = \xi_{1}'\mathbf{x}_k( \mathbf{x}_k'\mathbf{w} - x_{1k}) = \xi_{1}'\mathbf{Z}_k( \mathbf{Z}_k'\mathbf{w} - Z_{1k})$. Recall the separation assumption,
$$
\bm{w}^* \in \text{argmin}_{w\in \Delta^J} \mathbb{E} \|\bm Y^{val}_1 - \bm Y_0^{val}\bm w\|^2
$$ 
\noindent satisfies that for all $k\in S$ and $l \in S^c$, $|Z_{1k} - Z_{Jk}'\bm{w}^*| > 0$. By the reverse triangle inequality
\begin{align*}
    |Z_{1k} - Z_{Jk}'\bm{w}| &= |Z_{1k} - Z_{Jk}'\bm{w}^* - Z_{Jk}'(\bm{w} - \bm{w}^*)| \\
    &\geq \left| |Z_{1k} - Z_{Jk}'\bm{w}^* | - |Z_{Jk}'(\bm{w} - \bm{w}^*)| \right| \\
    &\geq  |Z_{1k} - Z_{Jk}'\bm{w}^* | - \sqrt{k}\sum_j (w_j - w^*_j) .
\end{align*}

\noindent given that by assumption we have that $\|Z_j\|^2 \leq \sqrt{k}$. It follows that if for fixed $J$ $\bm{w} \overset{p}{\to} \bm{w}^*$ as $T_{val} \to \infty$ then $P(|Z_{1k} - Z_{Jk}'\bm{w}| > 0 ) \to 1$ as $T_{val} \to \infty$ so we don't have convergence in probability. Furthermore, under the injective assumption of $\psi$ the $Z$ are linearly independent so it follows that $\xi_{1}'\mathbf{Z}_k \neq 0$ as by feasibility and the weight sum to one restriction $\xi_{1j}$ can not be zero for all $j$. Therefore, it follows that for fixed J, if as $T_0, T_{val} \to \infty$, $\bm{w} \overset{p}{\to} \bm{w}^*$ and $\lambda \to 0$, then
$$
P(\rho_k = 0) \to 0.
$$
\noindent This implies that $P(\kappa_k^v = 0) \to 0$ and, therefore, that for $k\in S$ for fixed J, if as $T_0, T_{val} \to \infty$, $\bm{w} \overset{p}{\to} \bm{w}^*$ and $\lambda \to 0$, then
$$
P(v_k =0) \to 1.
$$
\noindent This shows the first part of the theorem. If, furthermore, we have that for $l \in S^c$ $|Z_{1k} - Z_{Jk}'\bm{w}^* | = 0$ (i.e. perfect fit) then a similar argument yields that for fixed J, if as $T_0, T_{val} \to \infty$, $\bm{w} \overset{p}{\to} \bm{w}^*$ and $\lambda \to 0$, then
$$
P(v_l = 0) \to 0.
$$
Next, we show that for fixed $J$ we have that $\bm w \overset{p}{\to} \bm w^*$ as $T_0, T_val \to \infty$. Let $Q = \mathbb{E} \|\bm Y^{val}_1 - \bm Y_0^{val}\bm w\|^2$ and $\hat{Q} = \frac{1}{T_{val}} \|\bm Y^{val}_1 - \bm Y_0^{val}\bm w\|^2$. Given the covariance stationarity assumption by a suitable LLN it follows that for any $w$ we have that $\tilde{Q} \overset{p}{\to} Q$ as $T_{val} \to \infty$. Hence, it follows that $\sup_w |\tilde{Q}(\bm w) - Q(\bm w)| \to 0$ as $T_{val} \to \infty$. Furthermore, we have that $\tilde{Q}$ and $Q$ are continuous and convex, and the simplex $\Delta^J$ is compact. Therefore, by Lemma 3 of \textcite{amemiya1973} (or Theorem 2.1 in \textcite{newey}) it follows that $\bm\tilde{w} \overset{p}{\to} \bm w^*$ where 
$\bm \tilde{w} \in \text{argmin}_{w\in \Delta^J} \hat{Q}(\bm w)$.

In general $\bm \tilde{w}$ need not be equivalent to the $\bm w$ that solves the bi-level program. However, as $T_0, T_{val} \to \infty$ and $\lambda$ is chosen to minimize the validation loss as in our bilevel program, $\bm \tilde{w} \overset{p}{\to} \tilde{w}$. The main observation is that the bilevel loss nests the $\tilde{Q}$ loss at the optimum. Indeed, set all $v_k$ weights to $0$ and $\lambda = 0$, then the lower level problem loss depends on $|Z_{1k_0} - Z_{Jk_0}'w|$ and the upper level problem loss is equivalent to $\tilde{Q}$. Under the oracle separation assumption it follows that we only need to consider the upper level loss as it will simultaneously minimize the lower level problem. For instance, if we can perfectly fit the predictors $k\in S^c$ it follows that as $T_0, T_{val} \to \infty$ our bilevel loss coincides with $\tilde{Q}$ therefore it follows by the trivial inequality (can't do better than $\bm w^*$) that $\bm w \overset{p}{\to} \bm w^*$.

\subsection{Proof of Theorem 2: MSE Rates}

We focus on the covariate matching problem with subGaussian noise:

$$
Z_1 = Z_0w^* + \epsilon, \quad \epsilon \sim_{ind} \text{subG}(\sigma_z^2).
$$

Under A1-A3 we assume a sparse representation where only $k_1$ predictors are non-zero. Theorem 1 implies that as $T_0 \to \infty$ almost surely the sparse predictors are not used in the model. Hence, with probability one, we can re-write the optimization problem in terms of the true underlying sparse model:

\begin{align*}
	min_{V,W} &L_V(V,W) = \frac{1}{T_{val}} \|Y_1^{val} - Y_0^{val}W(V)\|^2 + \lambda \|V\|_1, \\
	\text{s.t.  } &W(V) \in \psi(V),\\
	& V \in \mathcal{V},
\end{align*}

\noindent
where $\psi : \mathcal{V} \rightrightarrows \mathcal{W}$ maps the upper level solutions to the lower level optima

$$
\psi(V) \equiv argmin_{W\in \mathcal{W}} L_W(V,W) = \|X^{train}_1 - X^{train}_0W\|^2_V,
$$

\noindent and
\begin{align*}
	W\in \mathcal{W} &\equiv \left\{W\in \mathbb{R}^J\text{ } |\text{ } \mathbf{1}'W = 1, \text{ } W_j\geq 0, \text{ } j=2,\dots, J+1\right\} \equiv \Delta^J, \\
	V\in \mathcal{V} &\equiv \{ (v_1, \dots, v_k) \mid v_j = 0 \text{ for } j\in S, \text { } v_j\in \mathbb{R} \text{ for }  j=1,\dots k\}.
\end{align*}
\noindent where $S = \{k \mid \theta_{tk} = 0 \text{ for all } t\}$ denotes the set of nuissance predictors, and $|S^c| = k_1$. Consider the lower level program of matching the covariates. For simplicity, restrict design matrix to the covariates, without including transformations of the outcome variable. Furthermore, let $Z$ be bounded such that $\max_{j} \|Z_j\| \leq \sqrt{k}$ as in A1-A3. Denote the minimizer of the lower level program by $\hat{w}$, then it follows that
\begin{align*}
	\| Z_1 - Z_0\hat{w}\|^2_V &\leq \| Z_1 - Z_0w^*\|^2_V \\
	\| Z_0w^* - Z_0\hat{w}\|^2_V &\leq 4 \langle V^T \epsilon, \frac{Z_0w^* - Z_0\hat{w}}{\|Z_0w^* - Z_0\hat{w}\|_V}\rangle^2 \\
	&\leq \sup_{b\in Z_0 \Delta^J, \|b\|=1} 4 \langle V^T \epsilon, b\rangle^2
\end{align*}
\noindent Given our assumptions it can be shown that $Z_j^T V^T \epsilon \sim \text{subG}(k_1 \sigma_z^2)$. Therefore, using a maximal inequality it follows that
$$
MSE(Z_0\hat{w}) = \frac{1}{k} \mathbb{E} \max_b \langle V^T\epsilon, b\rangle \lesssim \frac{\sigma_z \sqrt{k_1}}{k} \sqrt{2 \log J}.
$$

\noindent The rate for the standard synthetic control can be derived in a similar way, without restricting the model to be in the non-sparse support of the predictors.

\subsection{Placebo variance estimation}

We estimate the noise level using the placebo variance estimation bootstrap proposed in \textcite{sdid}, which follows from the placebo exercises first described by \textcite{adh2010}. This bootstrap procedure is valid when the noise distribution of the treated and donor units is the same (homoskedasticity). A further discussion on similar bootstrap procedures when the number of treated units is small can be found in Conley and Taber (2011).

\noindent The bootstrap procedure works as follows. For $B$ replications:

\begin{enumerate}
    \item Randomly choose a donor unit $l$ to be treated.
    \item Compute the synthetic control weights $\hat{\bm w}^b$ using the remaining controls units as the donor pool. 
    \item Estimate the average treatment effect on the treated in the post treatment period
    $$
    \hat{\tau}_b = \frac{1}{T-T_0} \sum_t Y_{lt} - \bm Y_{-lt}'\hat{\bm w}^b.
    $$
\end{enumerate}

\noindent Then the placebo variance estimator is given by
$$
\hat{V}_{\tau} = \frac{1}{B} \sum_{b} (\hat{\tau}_b - \bar{\tau}_b)^2,
$$
\noindent where $ \bar{\tau}_b = \frac{1}{B} \sum_b \hat{\tau}_b$. In a simulation exercise, \textcite{sdid} show that the coverage properties of the placebo variance bootstrap procedure are good.

\subsection{Additional Simulation Figures}

\begin{figure}
  \vspace{-3cm}
   \begin{subfigure}[b]{0.5\textwidth}
    \vspace{-2.5cm}
    \hspace*{-1cm}
  	\includegraphics[width=1.2\linewidth]{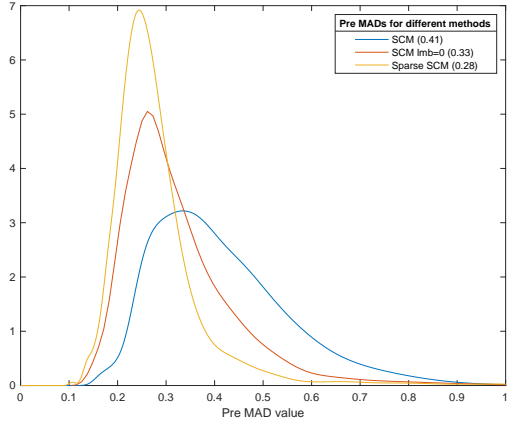}
  	\vspace*{-3.5cm}
  	\caption*{(c) $k_1 = k_2 = 5.$}
  \end{subfigure}
  \begin{subfigure}[b]{0.5\textwidth}
    \vspace{-2.5cm}
  	\hspace*{-1cm}
  	\includegraphics[width=1.2\linewidth]{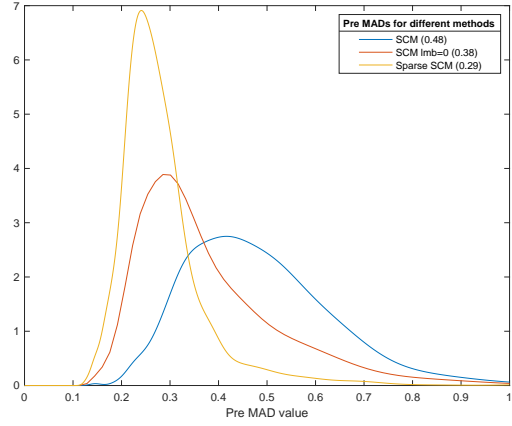}
  	\vspace*{-3.5cm}
  	\caption*{(d) $k_1 =1$, $k_2 = 9.$}
  \end{subfigure}
  \hspace*{-1.1cm}
  \begin{subfigure}[b]{0.5\textwidth}
     \vspace{-2.5cm}
  	\includegraphics[width=1.2\linewidth]{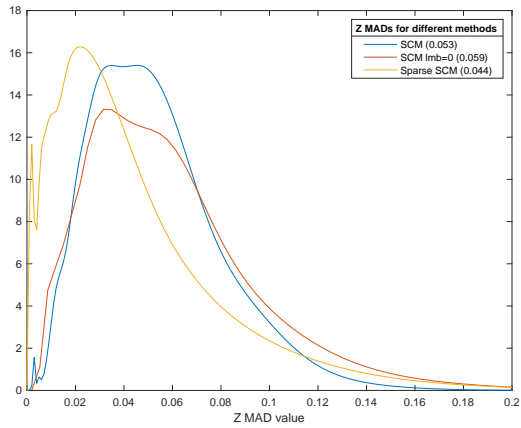}
  	\vspace*{-3.5cm}
  	\caption*{ \hspace*{1.5cm} (e) SCM $\lambda^* = 0$ $\bm V^*$}
  \end{subfigure}
  \begin{subfigure}[b]{0.5\textwidth}
    \vspace{-2.5cm}
  	\includegraphics[width=1.2\linewidth]{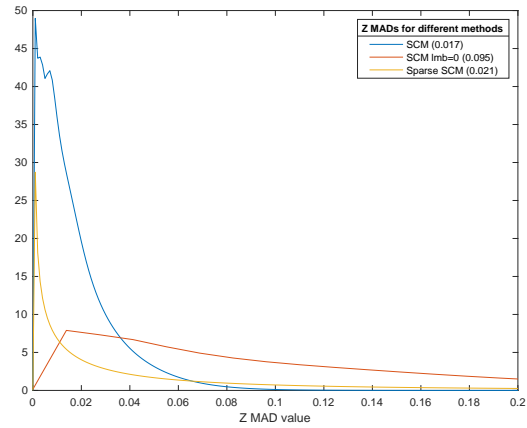}
  	\vspace*{-3.5cm}
  	\caption*{ \hspace*{1.5cm} (f) \textit{Sparse} SCM $\bm V^*$}
  \end{subfigure}
  \label{fig:mads}
  \caption{Pre-treatment Fit and Predictor Fit.}
  \begin{tablenotes}
		\small \item \textbf{Notes}: Panels (a) - (b) show the kernel density across simulations of MADs of the outcome variable for the pre-treatment period, with average values in parenthesis. Panels (c) - (d) show the kernel density across simulations of MADs of the useful predictors, with average values in parenthesis. 
   \end{tablenotes}
\end{figure}

\end{document}